\documentclass[
reprint,
superscriptaddress,
showpacs,
 amsmath,amssymb,
floatfix,
]{revtex4-1}
\usepackage{dcolumn}
\usepackage{bm}
\usepackage{turnstile}

\usepackage{graphicx}
\usepackage{amsmath,amssymb}
\usepackage{url}

\usepackage{epsfig}
\usepackage{epstopdf}

\begin{document}

\title{
Periodic Co/Nb pseudo spin-valve for cryogenic memory
}
\author{N.~V.~Klenov}
\affiliation{Skobeltsyn Institute of Nuclear Physics, Moscow State University, Moscow 119991, Russia}
\affiliation{Moscow Institute of Physics and Technology, Dolgoprudny, Moscow Region,
141700, Russian Federation}
\affiliation{All-Russian Research Institute of Automatics n.a. N.L. Dukhov (VNIIA), 127055, Moscow, Russia}

\author{Yu.~N.~Khaydukov}
\email [Electronic mail: ]{y.khaydukov@fkf.mpg.de}
\affiliation{Max-Planck-Institut f\"ur Festk\"orperforschung, Heisenbergstra\ss e 1, D-70569 Stuttgart, Germany}
\affiliation{Max Planck Society Outstation at the Heinz Maier-Leibnitz Zentrum (MLZ), D-85748 Garching, Germany}
\affiliation{Skobeltsyn Institute of Nuclear Physics, Moscow State University, Moscow 119991, Russia}

\author{S.~V.~Bakurskiy}
\affiliation{Skobeltsyn Institute of Nuclear Physics, Moscow State University, Moscow 119991, Russia}
\affiliation{Moscow Institute of Physics and Technology, Dolgoprudny, Moscow Region,
141700, Russian Federation}
\author{I.~I.~Soloviev}
\affiliation{Skobeltsyn Institute of Nuclear Physics, Moscow State University, Moscow 119991, Russia}
\affiliation{Moscow Institute of Physics and Technology, Dolgoprudny, Moscow Region,
141700, Russian Federation}
\author{R.~Morari}
\affiliation{Institute of Electronic Engineering and Nanotechnologies ASM, MD2028 Kishinev, Moldova}
\author{T.~Keller}
\affiliation{Max-Planck-Institut f\"ur Festk\"orperforschung, Heisenbergstra\ss e 1, D-70569 Stuttgart, Germany}
\affiliation{Max Planck Society Outstation at the Heinz Maier-Leibnitz Zentrum (MLZ), D-85748 Garching, Germany}

\author{M.~Yu.~Kupriyanov}
\affiliation{Skobeltsyn Institute of Nuclear Physics, Moscow State University, Moscow 119991, Russia}
\affiliation{Moscow Institute of Physics and Technology, Dolgoprudny, Moscow Region,
141700, Russian Federation}

\author{A.~S.~Sidorenko}
\affiliation{Institute of Electronic Engineering and Nanotechnologies ASM, MD2028 Kishinev, Moldova}

\author{B.~Keimer}
\affiliation{Max-Planck-Institut f\"ur Festk\"orperforschung, Heisenbergstra\ss e 1, D-70569 Stuttgart, Germany}
\date{\today}

\begin{abstract}
We present a new study of magnetic structures with controllable effective exchange energy for Josephson switches and memory. As a basis for a weak link we propose to use a periodic structure comprised of ferromagnetic (F) layers spaced by thin superconductors (s). Our calculations based on Usadel equations show that switching from parallel (P) to antiparallel (AP) alignment of neighboring F layers can lead to a significant enhancement of the critical current through the junction. To control magnetic alignment we propose to use periodic system where unit cell is a pseudo spin-valve $F_1$/s/$F_2$/s with $F_1$ and $F_2$ two magnetic layers having different coercive fields. In order to check feasibility of controllable switching between  AP and P states through the \emph{whole} periodic structure we prepared a superlattice [Co(1.5nm)/Nb(8nm)/Co(2.5nm)/Nb(8nm)]$_6$ between two superconducting layers of Nb(25nm). Neutron scattering showed that parallel and antiparallel alignment can be organized by using of magnetic fields of only several tens of Oersted.
\end{abstract}


\maketitle

Superconductor digital devices attract growing attention due to unique energy efficiency and performance, and also due to compatibility with a number of quantum computers under development \cite{soloviev2017beyond}. However the lack of cryogenic memory elements with fast enough switching between stable states, small enough energy dissipation in the process, is still the main obstacle in the field. A use of competition and coexistence of superconducting (S) and ferromagnetic (F) correlations promises an increase in the degree of integration for cryogenic memory storage devices \cite{soloviev2017beyond,ryazanov2012,goldobin2013,baek2014,Alidoust14,golod2015,Bakurskiy2016, Shafranjuk16,gingrich2016,NevirkovetsPRA18}. Some of these interesting implementations requires a Josephson contact with two stable states: a high value of the critical current, $I_C$, corresponds to the "open" state and a low value - to the "closed" state. Such a device can be organized if a weak link is a composite F/N/F trilayer (N is a normal metal) which magnetic state can be switched between parallel and antiparallel directions of magnetization vector of the F layers \cite{baek2014}.

\begin{figure*}[htb]
\centering
\includegraphics[width=2\columnwidth]{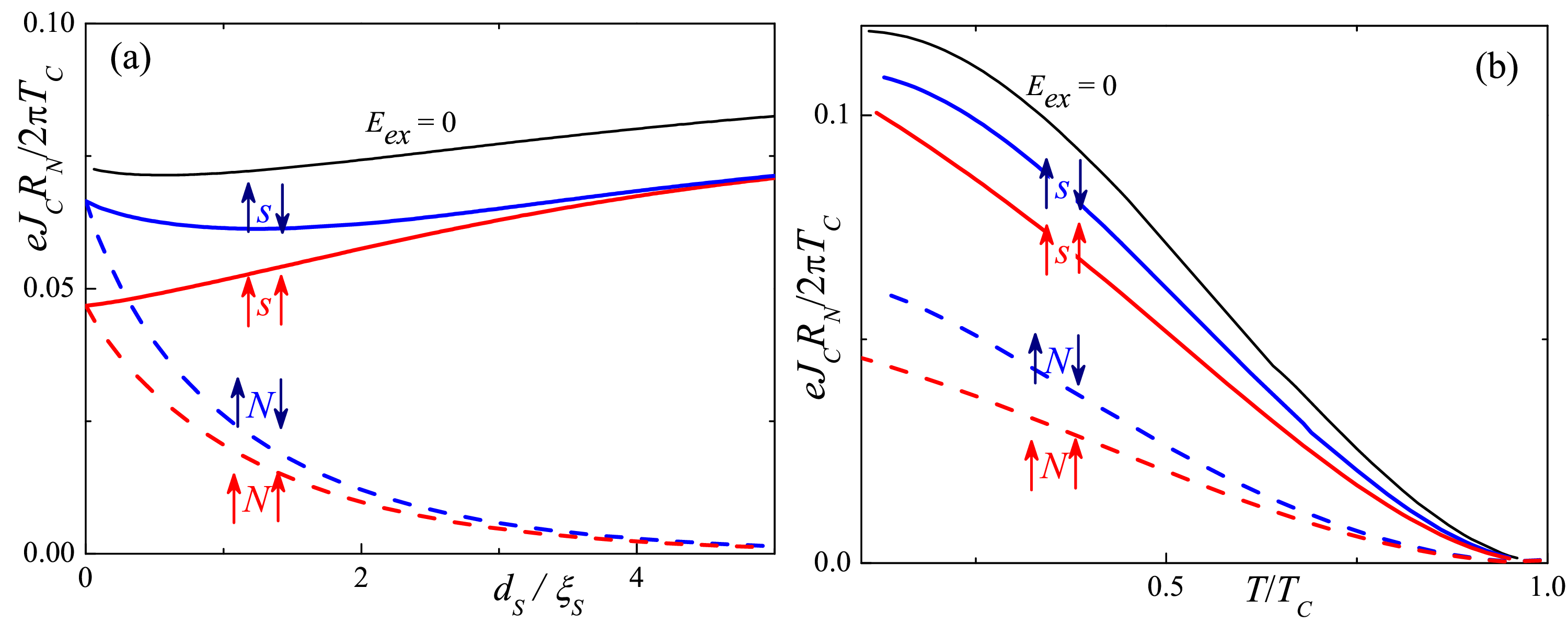}
\caption{
The normalized critical current of the S/F/s/F/S (solid lines) and S/F/N/F/S (dashed lines) structures as a function of the thickness of the  spacing layer (a) and the dimensionless magnitude of the temperature (b).  The red lines correspond to the case when the exchange energy in both layers of the ferromagnet is equal in magnitude, and the magnetization vectors lying in the plane of the magnetic layers are parallel (P). The blue lines are for the case when the exchange energy, $E_{ex}$, in both F-layers is equal, and the described magnetization vectors are antiparallel (AP). The black curves correspond to the $E_{ex}$ =0 case.
}
\label{Fig1}
\end{figure*}

Use of  a thin superconducting layer (s) as a spacer instead of a N layer may lead to enhancement of the spin valve effect due to proximity of the thick superconductor banks and the thin  superconductor spacers (see, e.g. \cite{Alidoust14}). To check this hypothesis we calculated critical current of S/F/s/F/S and S/F/N/F/S Josephson junctions (Fig. \ref{Fig1}). The calculations were performed in the frame of Usadel equations \cite{Usadel} with Kupriyanov-Lukichev conditions \cite{KL} for parallel (P) and antiparallel (AP) orientations of F films magnetization vector, $\textbf{M}$. Fig. \ref{Fig1}(a) and (b) show $I_C$ dependence on reduced thickness of a spacer, $d_s/ \xi_S$, and temperature, $T/T_C$. Here $T_C$ is the critical temperature of the S electrodes and the  s layer  material, $ \xi_S = (D_S/2 \pi T_C) ^{1/2}$ is a superconducting correlation length and $D_S$ is a diffusion coefficient. The calculations were performed for thickness of the F films $d_{F1}=0.1\xi_F$ and $d_{F2}=0.15\xi_F$, suppression parameters at the S/F and s/F interfaces, $\gamma_B =R \mathcal{A}/ \xi_S \rho_S = 0.3,$ and exchange energy, $E_{ex} =10 T_C.$ For simplicity, we assumed that $\xi_S=\xi_F$ and $\rho_S=\rho_F$.

As it follows from Fig. \ref{Fig1} the existence of intrinsic superconductivity of the spacer significantly increases $I_C$  of the S/F/s/F/S comparing to S/F/N/F/S junction. The effect can be essentially enhanced in S/[F/s]$_n$/F/S Josephson devices with the superlattice in the weak link region. The use of a multilayer structure has several advantages. In it, thanks to the collective effect of maintaining the superconducting state in spacers, it is possible to use more thinner s  layers. The thinning of the layers should be accompanied by a decrease in effective exchange energy  due to its renormalization
\cite{PhysRevLett.86.3140,PhysRevB.66.014507}.  Moreover, for the AP orientation of the magnetization vectors of the F  layers, another additional mechanism arises for the renormalization of the effective exchange energy, 
 which leads to its further decrease \cite{PhysRevB.69.024525,PhysRevLett.109.237006,Bakurskiy2015}.

\begin{figure*}[htb]
\centering
\includegraphics[width=2\columnwidth]{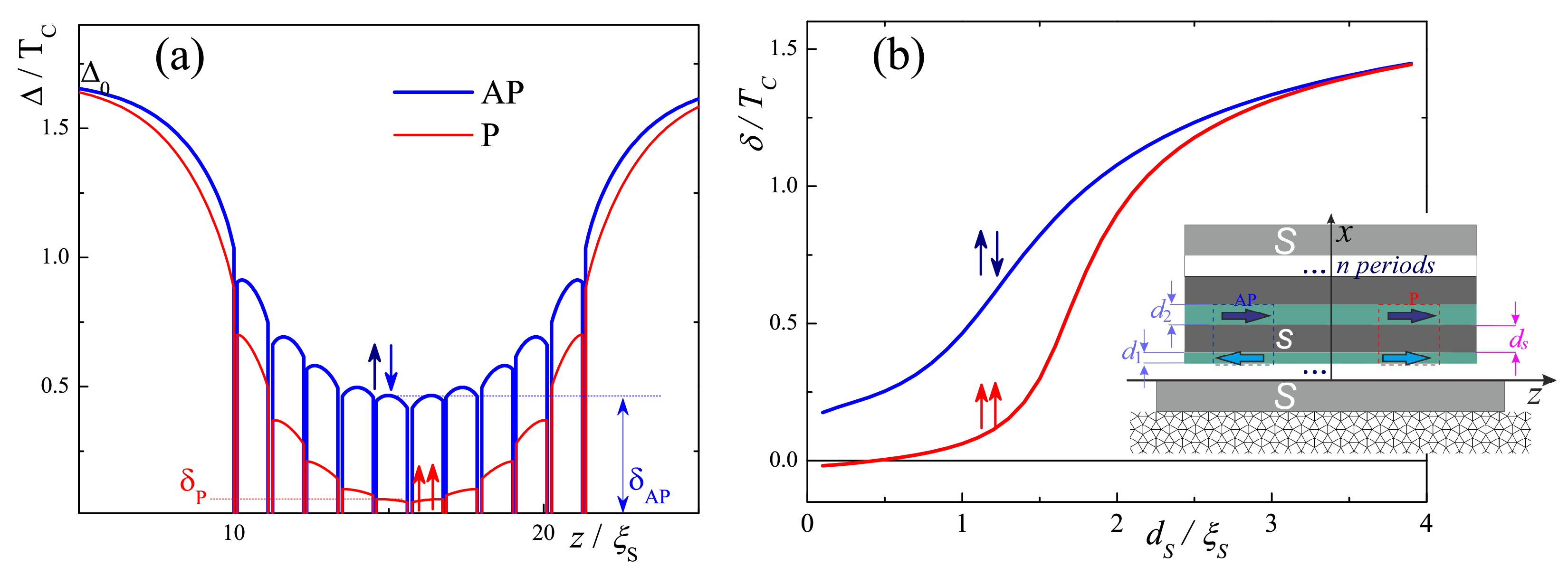}
\caption{
(a) The depth profile of the superconducting order parameter amplitude of  $S/[F_1/s/F_2/s]_5/F_1/S$ structure in the P and AP cases. (b) The amplitudes of the superconducting order parameter in the middle of the weak link for the same situations. Inset - schematic representation of the considered stack structure.
}
\label{Fig2}
\end{figure*}

To prove these statements, we have generalized the  S/[F/N]$_n$ model \cite{Bakurskiy2015}
to the case of the existence of intrinsic superconductivity in its non-ferromagnetic parts. To make the model more realistic we consider a case of a periodic pseudo spin-valve structure. In it two neighboring F layers have slightly different thicknesses $d_{F1}$ and $d_{F2}$ (see inset in Fig. \ref{Fig2}b). The difference in the thicknesses of the F$_1$ and F$_2$ layers provides a difference of their coercive fields $H_{c1} \neq H_{c2}$ which allows to organize AP state in the range of magnetic fields, $H,$ max($H_{c1},H_{c2}$) \textgreater $H$ \textgreater min($H_{c1},H_{c2}$) after saturation of the layers magnetizations in negative direction. Figure \ref{Fig2}(a) shows the spatial distribution of the order parameter amplitudes in the S/[F$_1$/s/F$_2$/s ]$_n$/F$_1$/S structure for the P and AP alignments. The calculations were performed for the same set of parameters as in  Fig.\ref{Fig1}. From  Fig. \ref{Fig2}a it follows that the considered structure is a series connection of s/F$_1/s$ and s/F$_2/s$ Josephson junctions with the weakest place located in the middle of the structure. Figure \ref{Fig2}(b) shows the amplitudes of the order parameter, $\delta_P,$ and  $\delta_{AP},$ (see the definition  of  $\delta_P$ and  $\delta_{AP}$ in  Fig.\ref{Fig2}a) in the middle of the weak link as a function of s layers thickness. One can see that the amplitudes for AP and P configurations are significantly different for $d_s \sim \xi_S$. As soon as $I_C$ is proportional to the product of  order parameters amplitude of the s banks, one may estimate that the ratio of  $I_C$ for AP orientation to that for P one is of the order of $(\delta_{AP}/\delta_{P})^2 \approx 25.$ From  Fig. \ref{Fig2}b it follows that such the enhancement depends on the ration $d_s/ \xi_S$ and achieves maximum in the vicinity of $d_s=\xi_S$.

Realization of the  proposed above S/[F$_1$/s/F$_2/$s]$_n$/F$_1$/S Josephson devices  requires the development of a technology for manufacturing of multilayer structures that satisfy the following conditions: (a) presence of superconductivity in the s  layers with $T_C\gtrsim  4.2$K, (b) in plane orientation of magnetization vector in the F films, (c) ability for coherent switching between P and AP configurations through the whole stack. The goal of this paper is to demonstrate that the requirements can be met when using combination of Nb and Co as materials for the superlattice. To do this we fabricate Nb(25nm)/[Co(1.5nm)/Nb(8nm)/Co(2.5nm)/Nb(8nm)]$_6$ /Co(1.5nm)/Nb(25nm) structure. We take niobium  as a superconducting material since it has the highest $T_C$ = 9.25 K among all elemental superconductors and forms stable structures with cobalt\cite{obi1999oscillation,Stamopoulos2014,Liu2016,Lee2003,robinson2007,ObiPRL2005}. Thickness of the Nb-spacer was chosen to be close to $\xi_S \approx 6-10$nm, the value have been  found in our prior studies \cite{Zdravkov2010,Zdravkov2006}.   Thickness of the Co layers were in the range of $\xi_F \sim $ 1 nm \cite{ObiPRL2005}, which is enough to form homogeneous and magnetic layer \cite{obi1999oscillation}.


The sample was prepared using Leybold Z-400 magnetron machine at room temperature on a R-plane oriented sapphire ($Al_2O_3$) substrate. Before the deposition the  substrates were etched by argon ion beam inside the chamber. Targets of Nb($99.99\%$) and Co($99.99 \%$) were presputtered for cleaning from metalic oxides and contaminations absorbed by surfaces. Additionally, immediately before deposition of the next layer we presputtered the corresponding target during 40-50 seconds for stabilisation of the film growth rate. The deposition was performed in a pure argon atmosphere ($99.999\%$ purity) at working pressure of 8$\times$ 10$^{-3}$ mBar. The thickness of the films was controlled by the time of deposition of the material on the substrate. For high repeatability of thicknesses of thin Nb films, an electrical motor was used to move the target above the substrate at an equal speed thus, the thickness of the niobium layer remains the same for each of the periods of the structure. The growth film rate is 1 nm/s and 0.1 nm/s for Nb and Co respectively. After the deposition the structure was capped by a silicon layer.

\begin{figure*}[htb]
\centering
\includegraphics[width=2\columnwidth]{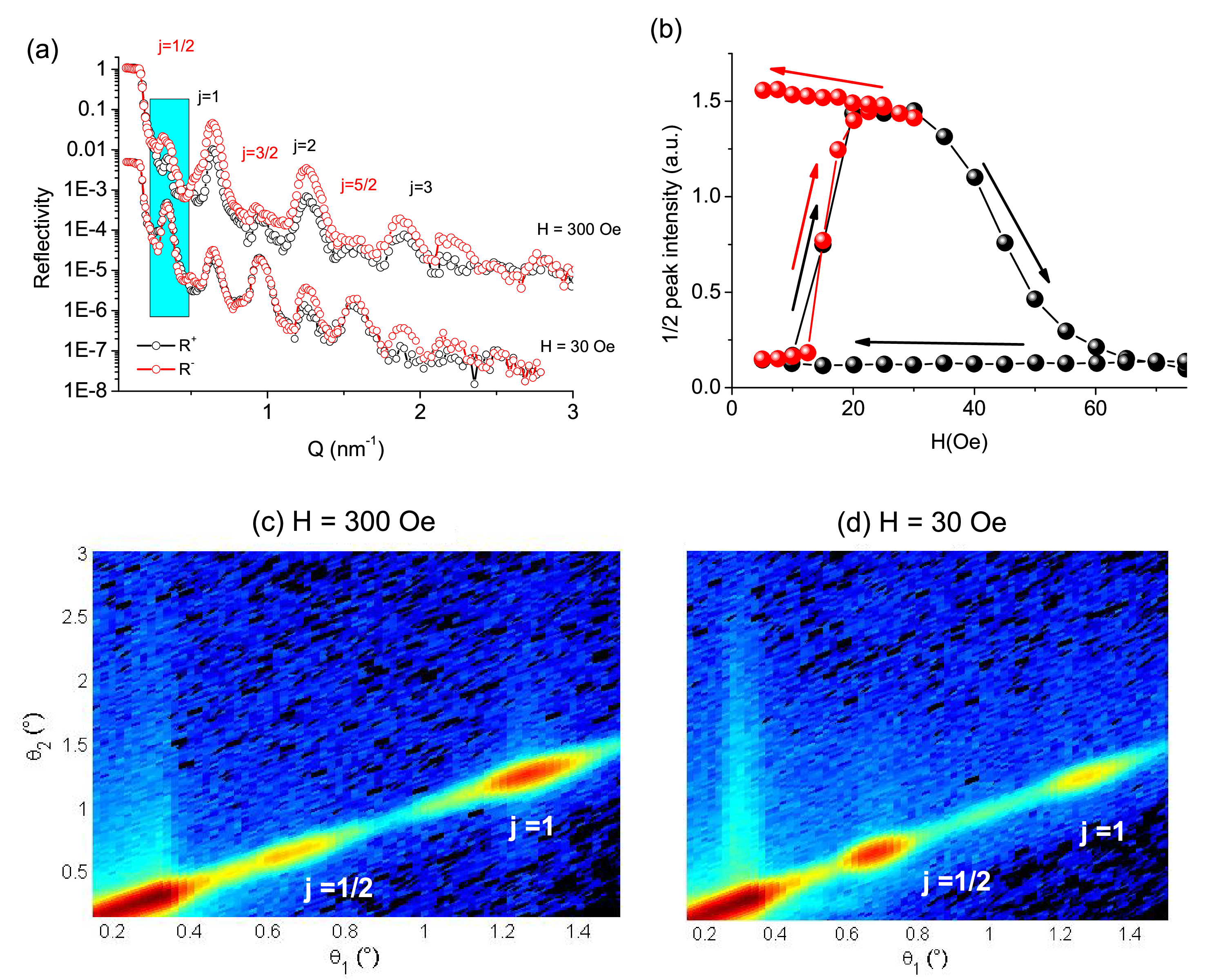}
\caption{
(a) specular neutron reflectivities measured at $T$=13K. Numbers above shows the corresponding order of Bragg reflection from effective [Co(2nm)/Nb(8nm)]$\times$12 periodic structure. Blue rectangular shows the area of 1/2 peak which field dependence is shown in (b). The scattering maps of spin-down neutrons measured at $H$ =300 Oe and $H$ =30 Oe are shown in (c) and (d).
}
\label{Fig3}
\end{figure*}

In order to characterize structural and magnetic ordering of the Co/Nb superlattice we have performed neutron small angle diffractometry (reflectometry). Measurements were conducted at the neutron reflectometer NREX at the research reactor FRM-2 (Munich, Germany). Neutron reflectivities were taken with a monochromatic polarized neutron beam with  wavelength $\lambda$ =0.43 nm at a temperature $T$=13K with magnetic field applied in-plane of the structure and normal to the scattering plane.
Fig.\ref{Fig3}(a) shows reflectivities measured at saturation ($H$=300 Oe) and in magnetic field $H$ = 30 Oe after magnetization of the sample in negative magnetic field of $H$=-4.5 kOe. The curves at saturated state are characterized by the strong peaks at the positions $Q_j \approx j\times$ 0.63 nm$^{-1}$ (j=1,2,3). These peaks correspond to the Bragg reflection from the period $D = 2\pi/Q_1 \approx$ 10 nm. This means we can effectively consider our structure as [Co(2nm)/Nb(8nm)]$_{12}$ with $\sim$ 20 \% variation of the F layer thickness. The reflectivity pattern at  $H$ = 30 Oe strongly differs from the saturated state. First of all we can see the appearance of non-integer peaks $j$/2 which directly evidence the doubling of magnetic period at this field \cite{Nagy2002,Lauter-Pasyuk2002,Langridge2000,rehm1997,Hjoervarsson1997}. The small difference of $R^+$ and $R^-$ peaks speak about compensation of magnetic moments of neighboring Co layers, e.g. antiparallel alignment. Figure \ref{Fig3}(b) shows the field evolution of the 1/2 peak. One can see that the AP alignment exists in the range of magnetic fields $H$ = (10-30) Oe if the sample was firstly magnetized in negative direction. Moreover once the AP state is created the field can be released to zero and the alignment will be saved. The P-alignment can be also organized at zero field if the sample was saturated before in positive field.

An important question is the lateral homogeneity of the obtained magnetic configurations. So the systems with exchange coupling of F layers through N spacers are characterized by laterally non-homogeneous AP state. This domain state can be directly evidenced in neutron experiment as strong off-specular scattering in the vicinity of 1/2 peak \cite{Nagy2002,Lauter-Pasyuk2002,Langridge2000}. In our case we have not observed strong increase of the diffuse scattering intensity in the AP state (Fig.\ref{Fig3}(d)) comparing to the saturation (Fig.\ref{Fig3}(c)).

\begin{figure}[htb]
\centering
\includegraphics[width=1\columnwidth]{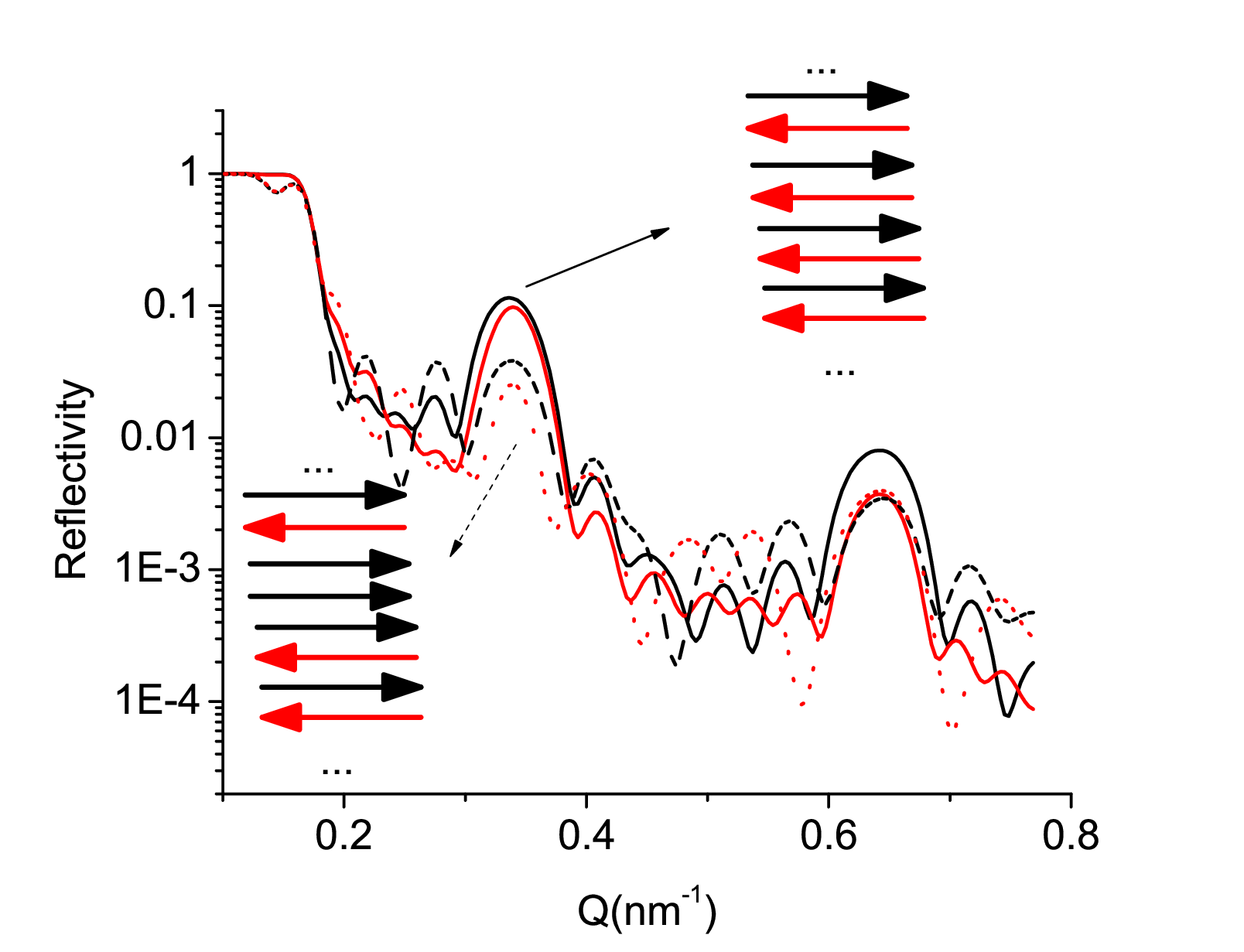}
\caption{
Model $R^+$ (black) and $R^-$ (red) reflectivity curves for the AP configuration. Solid lines show reflectivity curves for ideally AP aligned structure (see right inset) while dashed lines are for the structure with stacking fault in the middle of the structure (see left inset).
}
\label{Fig4}
\end{figure}

In a periodic pseudo spin-valve structures one can not exclude the situation of non coherent switching of the F layers. Such a stacking fault in the antiferromagnetically aligned system may lead to the suppression or even destruction of the spin-valve effect. Our simulations show (Fig.\ref{Fig4}) that presence of two ferromagnetically aligned neighboring Co layers in the middle of AP aligned lattice would lead to the significant suppression of the 1/2 peak down to $R(Q_{1/2}) \sim$ 2 - 4\% and appearance of its polarization ($R^+ \not=  R^-$). On the experiment, however, we saw much stronger 1/2 peak of $R(Q_{1/2}) \sim$ 10\% intensity having almost no polarization, proving thus high coherence of the AP alignment through the whole stack.

In conclusion, we propose here to use as a memory element a Josephson junction where a weak link is comprised of periodic S/F structure which can be switched between AP and P states. In the frame of  Usadel  equations we show that critical current across the junction significantly depends on the magnetic state of the periodic structure. In order to switch between AP and P states we propose to use periodically repeated quadrolayer $F_1/s/F_2/s$ where magnetic layers $F_1$ and $F_2$ have slightly different coercive field. In order to experimentally investigate the switching processes between P and AP states we sandwiched periodic structure [Co(1.5nm)/Nb(8nm)/Co(2.5nm)/Nb(8nm)]$\times$6/
Co(1.5nm)  between two Nb(25nm) electrodes. By the use of neutron reflectometry we have demonstrated that AP state can be organized and erased by applying field of 30 Oe.

We would like to thank V.L. Aksenov for the fruitful discussions. This work was supported by grant No. 18-72-10118 of the Russian Science Foundation. AS would like to thank the support of the project of the Moldova Republic National Program "Nonuniform superconductivity as the base for superconducting spintronics" ("SUPERSPIN", 2015-2018), grant STCU \#6329 (2018-2019) and the "SPINTECH" project of the HORIZON-2020 TWINNING program (2018-2020). YK would like to acknowledge DFG collaborative research center TRR 80. This work is based on experiments performed at the NREX instrument operated by Max-Planck Society at the Heinz Maier-Leibnitz Zentrum (MLZ), Garching, Germany.

\bibliography{CoNb_Refs}

\end{document}